\documentclass[12pt]{article}
\usepackage{sw20aip}

\input tcilatex
\begin{document}

\title{Pair formation in two electron correlated chains.}
\author{J. F. Weisz$^1$ and F. Claro$^2$ \\
\\
($^1$) INTEC (CONICET-UNL), Guemes 3450, 3000 Santa Fe, Rep. Argentina.\\
\\
($^2$) Pontificia Universidad Cat\'{o}lica de Chile, Santiago, Chile.}
\maketitle

\begin{abstract}
We study two correlated electrons in a nearest neighbour tight- binding
chain, with both on site and nearest neighbour interaction. Both the cases
of parallel and antiparallel spins are considered. In addition to the free
electron band for two electrons, there are correlated bands with positive or
negative energy, depending on wheather the interaction parameters are
repulsive or attractive. Electrons form bound states, with amplitudes that
decay exponentially with separation. Conditions for such states to be filled
at low temperatures are discussed.
\end{abstract}

Exact solutions to problems involving correlated motion of interacting
particles are extremely rare. Even simple systems like electrons moving in a
wire are usually solved approximately only. A body of recent literature
exists for the case of just two electrons moving in a one dimensional
disordered potential.[1-9] The problem of N particles in an ordered string
of length L is conceptually simple if one ignores spin, since it is then
formally equivalent to that of a single particle in N-dimensional space,
with the pair interaction acting as a defect potential associated to the
planes $x_i=x_j$, where $x_i$ is the position of the i-th particle. One then
expects a band of about $L^N$ extended states with finite amplitude in all
of space, and bands of $L^{(N-S)}$ surface states localized about the
geometrical defect where $S$ planes intersect, or the planes themselves if $%
S=1$. In spite of this qualitative understanding of the ordered case, an
exact solution has been reported for an N=2 singlet state only.[10]. Thus,
in this reference, as well as previous work[1-9] it is found that two
correlated electrons exhibit a density of states characteristic of one
particle in a plane, that then has a Van Hove singularity at $E=0$.

In this work we report on another exact solution of the two-particle problem
in an ordered lattice. It describes a paired triplet state, with an energy
that may fall within the conduction band, making it of interest to
superconductivity.[11] In order to see how this comes about, we consider a
chain of L sites within the tight binding model, with up to nearest neighbor
interaction. With the understanding that the amplitudes $c(l,m)$ represent
two electrons at sites $l$ and $m$ with either parallel or antiparallel
spins, the equation of motion without disorder reads,

$-tc(l-1,m)-tc(l+1,m)-tc(l,m+1)-tc(l,m-1)+$

$J[\delta _{l+1,m}+\delta _{l-1,m}]c(l,m)+U\delta _{l,m}c(l,m)=Ec(l,m)$%
\hspace{1.0in}(1)

A transformation to center of mass coordinates is effected by taking

$c(l,m)=\exp (ik(l+m)a)\chi (l-m).$\hspace{1.0in}(2)

Here $k$ is the center of mass momentum and $a$ is the lattice constant.
Denoting by $p=l-m$ the distance between the two electrons, the equation
obtained by substituting (2) into (1) is that of a single particle in an
effective linear chain with sites p,

$-2t\cos (ka)\chi (p-1)-2t\cos (ka)\chi (p+1)+J[\delta _{p,1}+\delta
_{p,-1}]\chi (p)+U\delta _{p,0}\chi (p)=E\chi (p)$\hspace{1.0in}(3)

where U is a contact Hubbard parameter[12] and J the nearest neighbor
coupling strength. A peculiar feature of this equation is that the effective
hopping parameter depends on $k$, actually vanishing at $ka=\pm \frac{\pi }{2%
}$. In the absence of interactions one may set $\chi (p)=\exp (ipqa)$,
obtaining the energy band

$E(k,q)=-4t\cos (ka)\cos (qa)$\hspace{1.0in}(4)

This covers the range $-4t<E<4t$. Assuming L large, Eq.(3) represents an a
lattice with impurities around the origin and we expect the band (4) to hold
true even in the interacting case, save for corrections in the density of
states of order 1/L. One can easily show the dispersion (4) to be exact in
the case $J=0$.[13]

We will first work out the magnetic case of two parallel spins. The spatial
wave function must then be antisymmetric under exchange of particles, or $%
\chi (p)=-\chi (-p)$. We consider equation (3) for the separate cases $p=0$, 
$p=1$ and $p\geq 2$, taking $\chi (0)=0$. The equation for $p=0$ is
trivially satisfied because of the antisymmetry of the wave function. In
looking for a solution for $p\geq 1$ we assume there is some constant $\beta 
$ such that $\chi (p+1)=\beta \chi (p)$. From the boundary condition that
the amplitudes must remain finite as $p$ becomes large, one must have $\mid
\beta \mid \leq 1$. Solving the equations we get

$E=J+(\frac{4t^2}J)\cos ^2(ka)$\hspace{1.0in}(5)

$\beta =-(\frac{2t}J)\cos (ka)$\hspace{1.0in}(6)

The solution decays exponentially with exponent $\gamma =\ln \mid J/(2t\cos
(ka))\mid $ , having the form $\chi (p)=A_{p}\exp (-\gamma \mid p\mid )$
where $A_{p}$ vanishes at the origin, and has the value $\frac{p}{\mid p\mid 
}$ if $\beta >0$ and $(-1)^{p}\frac{p}{\mid p\mid }$ if $\beta <0$, for
finite $p$. Since the amplitude is largest when the particles are next to
each other the state represents a pair bound state moving with center of
mass momentum $k$. Notice that in equations (5) and (6), $U$ is not involved
at all, a feature already contained in the Hubbard model[14]. Therefore it
was necessary to include nearest neighbor interaction to find it. Some
properties of this paired triplet state are the following. First, under a
change of sign of the nearest neighbor coupling constant, the energy just
changes sign. Referring now to positive $J$, at a fixed value of $k$ its
energy is above those of the free electron band at the same $k$. The state
may still be within such band, yet with different center of mass momentum.
The lowest energy is J and it occurs at $\mid ka\mid $ $=\pi /2$. In this
limit state the two electrons are as close as possible, with finite
amplitude as nearest neighbors only. For other values of $k$ the energy is
higher, and the pair is larger in size.

Band overlap between paired and free states occurs for $4t\geq J$. The
energy range of overlap is $J<E<2J$ if $2t\geq J$ and $J<E<4t$ otherwise.
The center of mass momentum is bounded by the condition $\mid \cos (ka)\mid
\leq \frac J{2t}$ in the former case, and $\mid \cos (ka)\mid \leq \sqrt{%
\frac Jt(1-\frac J{4t})}$ in the latter. If the Fermi energy lies above $J,$
when band overlap occurs, it will be energetically favorable to create bound
pairs, with wavector around $\mid ka\mid =\frac \pi 2$ , where the density
of states has a divergence. The model thus predicts an instability in the
Fermi liquid with the formation of pairs with parallel spin that coexist
with other free particle states. Since the pair and the single particle
momenta are not the same at a given energy, a transfer from one state to the
other requires some excitation (a phonon for example) to supply the missing
momentum.

Let us go on to consider the case of antiparallel spins. In this case the $U$
term is involved, and if ($U/J$) is large, we can ignore the nearest
neighbor coupling and solve the model with $J=0$, with $\chi (p)=\chi (-p)$.
Again we try $\chi (p+1)=\beta \chi (p)$ with $\mid \beta \mid \leq 1$.
Following a similar procedure as before we find now

$E=\sqrt{U^2+16t^2\cos ^2(ka)}$\hspace{1.0in}(7)

$\beta =\sqrt{\frac{E-U}{E+U}}$\hspace{1.0in}(8)

giving an exponential decay rate $\gamma $ $=\arcsin $h$|U/4t\cos (ka)|$.
The sign of the square root in equation (7) should be the same as the sign
of $U$ . This solution was already reported in Ref. 10. Its qualitative
features are similar to those of the previous one. Again, for positive
(negative) $U$ and same wavenumber $k$ the values given by equation (7) are
above (below) those given by equation (4). However, for $4t\geq U>0$,
allowing for different values of $k,$ there may be energies in the paired
band lower in energy than some in the single electron band. For positive $U$
the lowest energy state for the paired band is again at $ka=\pi /2$ and the
two electrons are stuck together, while as the energy increases above the
band minimum U the extent of the pair increases as well.

The most likely case in a real system is $U>\mid J\mid >0$. The paired
states with lowest energy will therefore correspond to parallel spins.
Different band widths for the paired states are predicted from equations (5)
and (7). Effects of overscreening[15] may possibly give rise to a positive $U
$, though a negative $J.$ As remarked before in the magnetic state $U$ does
not enter, and the energies for the paired solutions lie below the energies
of the single particle band center and, for $\mid J\mid $ sufficiently
large, even below the conduction band altogether.

In conclusion, the effect of correlations on interacting electrons, taken
two at a time, is to form paired states grouped in bands. In contrast to
various approaches based on Greens functions[17-18], our treatment is exact,
though with the limitation that we only consider correlations between two
electrons at a time. Correlations among more than two electrons may be
important. However, based on the equivalent picture of a single electron
moving in higher dimensions in a lattice with defects suggests that the
number of such states scales as $L^{-S}$. The possibility of having
unconventional metals has often been traced to the failure of Fermi liquid
theory in correlated lower dimensional systems.[19-22] It is obvious here
that the correlated bands cannot be placed in a one to one correspondence
with one electron states. Thus these effects may also be traced to the
possibility of two or more- particle correlations.

Acknowledgments

JFW is grateful for discussions with M.C.G. Passegui. FC thanks FONDECYT for
partial support under grant 1020829.

References

[1] D. L. Shepelyansky , Phys. Rev. Lett. \underline{73}, 2607 (1994)

[2] Y. Imry, Europhys. Lett. \underline{30}, 405 (1995)

[3] Dietmar Weinmann and Jean Louis Pichard, Phys. Rev. Lett. \underline{77}%
, 1556 (1996)

[4] Rudolph A. R\"{o}mmer and Michael Schreiber, Phys. Rev. Lett. \underline{%
78}, 515 (1997)

[5] S. N. Evangelou and D. E. Katsanos, Phys. Rev. \underline{B56}, 12797
(1997)

[6] P. H. Song and Doochul Kim, Phys. Rev. \underline{B56}, 12217 (1997)

[7] G. Benenti, X. Waintal and J.L. Pichard, Phys. Rev. Lett. \underline{83}%
, 1826 (1999)

[8] E. Gambetti-Cesare, D. Weinmann, R.A. Jalabert and P. Brune, Europhys.
Lett. 60, 120 (2002)

[9] Ph. Jacquod and D. L. Shepelyansky, Phys. Rev. Lett. \underline{78} 4986
(1997)

K. L. Frahm, A. Muller-Groeling, J. L. Pichard and D. Weinmann Phys. Rev.
Lett. \underline{75} 1598 (1995)

[10] F. Claro, J. F. Weisz and S. Curilef, Phys. Rev. B 67, 193101 (2003) 

[11] R. Micnas, J. Ranninger, S. Robaszkiewicz, Rev. of Modern Physics 
\underline{62}, 113 (1990)

[11] V. L. Ginzburg, Contemporary Physics \underline{33}, 15 (1992)

[12] J. Hubbard, Proc. Royal Society \underline{A276} 238 (1963)

[13] R. Feynman, R. Leighton and M. Sands, \underline{The Feyman lectures in
Physics}, Vol. III (1964)

[14] S. Doniach and E. H. Sondheimer, \underline{Greens Functions for Solid
State Physicists}, W. A. Benjamin Inc. (1974), Ch. 7

[15] J. E. Hirsch, D. J. Scalpino, Physical Rev. \underline{B32}, 5639 (1985)

[16] G. Rickayzen, \underline{Greens Functions and Condensed Matter},
Academic Press (1980)

[17] P. Fulde, \underline{Electron Correlations in Molecules and Solids},
Springer Berlin (1995), Third Edition

[18] W. Metzner, C. Castellani and C. di Castro, Advances in Physics 
\underline{47}, 3 (1998)

[19] K. V. Kusmartsev, J. F. Weisz, R. Kishore and M. Takahashi, Physical
Review \underline{B49}, 16234 (1994)

[20] See Ref. 16 above, and also J. M. Luttinger, J. Math. Phys. \underline{4%
}, 1154 (1963)

[21] F. D. M. Haldane, J. Physics \underline{C14}, 2585 (1981)

[22] A. Luther, Phys. Rev.\underline{ B14}, 320 (1979)

\end{document}